\documentstyle[preprint,aps,epsf]{revtex}                  
\global\let\epsfloaded=Y 
\begin{document}
\pagestyle{empty}                                      
\preprint{
\font\fortssbx=cmssbx10 scaled \magstep2
\hbox to \hsize{
\hfill $
\vtop{
 \hbox{ }}$
}
}
\draft
\vfill
\title{The $Z \to b \bar{b}$ Decay Asymmetry and Left-Right Models}
\vfill
\author{Xiao-Gang~He$^{(a)}$ and G.~Valencia$^{(b)}$}
\address{$^{(a)}$Department of Physics\\National Taiwan University\\
Taipei, Taiwan 10764, R.O.C.\\$^{(b)}$Department of Physics\\
Iowa State University\\Ames, IA 50011}

%
%
\vfill
\maketitle
\begin{abstract}

It has been pointed out recently by Chanowitz that the $Z \to b\bar{b}$ 
decay asymmetry poses a problem for the standard model whether it is 
genuine or not \cite{Chanowitz:2001bv}. If this conflict is interpreted 
as new physics in the $b$-quark couplings, it suggests a 
rather large right handed coupling of the $b$-quark to the Z-boson. 
We show that it is possible to  accommodate this
result in left-right models that single out the third generation.

\end{abstract}
%
%
%
%
\pagestyle{plain}

\section{Introduction}

The precision measurements at the $Z$ resonance continue to exhibit 
a deviation from the standard model in the observable $A^{b}_{FB}$ 
by about three standard deviations \cite{Drees:2001xw,Abbaneo:2001ix}. 
It has been pointed out recently by Chanowitz \cite{Chanowitz:2001bv} 
that this deviation indicates 
a problem whether it is genuine or not. In particular, Chanowitz argues 
that if the anomaly in $A^{b}_{FB}$ is attributed to systematic error 
and dropped from the LEP fits, then the indirect determination of the 
Higgs mass is in conflict with the direct limit \cite{Chanowitz:2001bv}. 

In Ref.~\cite{Altarelli:2001wx}, Altarelli {\it et.al.} approach this problem
by looking for super-symmetric corrections that improve the quality of the 
LEP fits (including $A^{b}_{FB}$), and that improve the consistency with 
the direct limits on the Higgs boson mass. They find that this is 
possible with light sneutrinos. 

The possibility of new physics affecting the $Zbb$ coupling has also 
been discussed in Ref.~\cite{Chanowitz:1999jj}. It is known that it 
is not easy to explain the  $A^{b}_{FB}$ anomaly with new physics 
in the $Zbb$ coupling mainly because the measurement of $R_b$ is in 
good agreement with the standard model. However, as pointed out 
by Chanowitz \cite{Chanowitz:1999jj}, it is possible to have deviations 
in both the left and right handed couplings of the $b$-quark to the $Z$-boson 
in such a way as to change  $A^{b}_{FB}$ without affecting  $R_b$. 

Our starting point is the combined fit to LEP and SLD measurements 
in terms of the left and right handed couplings of the $b$-quark. 
These are shown in Figure 11 of Drees~\cite{Drees:2001xw}, as well as 
in Ref.~\cite{Abbaneo:2001ix}. Subtracting the standard model values 
from the central value of the fit one obtains the deviations,
\begin{eqnarray}
\delta g_{Rb} &\approx & 0.02 \nonumber \\
\delta g_{Lb} &\approx & 0.001,
\label{fit}
\end{eqnarray}
where we have flipped the sign of $g_{Rb}$ in 
Ref.~\cite{Drees:2001xw,Abbaneo:2001ix} 
to agree with the particle data book \cite{Groom:in} 
definitions. 

The tree-level coupling in the standard model is written as  
\begin{eqnarray}
L(b_L) &=& -{g\over \cos\theta_W} \bar b \gamma^\mu (g_{Lb}L
+ g_{Rb} R) b Z_\mu,
\end{eqnarray}
with $L(R) = (1\mp\gamma_5)/2$. 
In terms of $g_{V} = t_{L3} - 2Q\sin^2\theta_W$, $g_A = t_{L3}$ 
(with the parameters defined in Ref.~\cite{Groom:in}), 
$g_{Lb} = (g_{Vb}+g_{Ab})/2$ and $g_{Rb} = (g_{Vb}-g_{Ab})/2$.
Here $t_{L3}$ is the weak isospin which is $1/2$ for up-type of quarks and
$-1/2$ for down-type of quarks, and $Q$ is the electric charge in units 
of $e$. At tree-level then,
\begin{eqnarray}
g_{Lb} &=& -{1\over 2} + {1\over 3} \sin^2\theta_W \, \sim -0.42 \nonumber \\
g_{Rb} &=&  {1\over 3} \sin^2\theta_W \, \sim 0.077
\end{eqnarray}
To gauge the magnitude of the required shifts, Eq.~\ref{fit}, 
it is useful to compare them with the one-loop correction in the standard 
model due to the heavy top-quark, 
$\delta g_{Lb} \sim 0.003$ \cite{Chanowitz:1999jj}. 

In view of the agreement of other low energy observables with the 
standard model, any new physics invoked to explain the 
$A^{b}_{FB}$ anomaly has to affect primarily the third family, and 
in particular the right-handed couplings.

Several scenarios in which the third generation interacts differently from 
the first two have been explored in the literature.  Foremost amongst these 
is top-color, where the $Zbb$ couplings have been studied extensively 
in connection with $R_b$~\cite{Burdman:1997pf,Yue:2000ay,Triantaphyllou:1998ke}. It is easy 
to see that while top-color can easily generate a correction to the 
left-handed $b$-quark coupling of the required magnitude, it cannot generate 
a sufficiently large correction to the right-handed $b$ coupling 
\cite{Yue:2000ay}. 
Other models considered in the literature, such as those of 
Refs.~\cite{Muller:1996qs},\cite{Malkawi:1999sa}, \cite{Malkawi:1996fs}
and~\cite{Lynch:2000md}, 
single out the third family as well. However, they 
predominantly affect the left-handed couplings, and cannot 
generate the shifts required by the $A^{b}_{FB}$ measurement. 
A possibility that may accommodate the required new physics 
appears in certain scenarios in which the $b$-quark mixes with heavy quarks 
with  unconventional charge assignments 
\cite{Bamert:1996px,Chang:1996pf,Chang:1998uj,Choudhury:2001hs}. 

Alternatively, the LEP data 
can be attributed to new physics in the form of higher dimension 
operators. In this way one does not have to explain the origin of 
the new physics but can still use it to predict other consequences. 
This has been done in Ref.~\cite{Oakes:1999zi}. 

In this paper we explore the possibility of a left-right 
model that preferentially affects the third family. In Section~2 
we present a model of this type and show how it can naturally 
accommodate the required shift in $g_{Rb}$. In Section~3 we 
explore the viability of the model in light of other existing 
constraints.

\section{The Model}

The specific model to be discussed is a variation of  
Left-Right models \cite{lrmodel,babu}.
The gauge group of the model is $SU(3)\times SU(2)_L\times SU(2)_R
\times U(1)_{B-L}$ with gauge 
couplings $g_3$, $g_L$, $g_R$ and $g$, respectively. 
The model differs from other Left-Right models in 
the transformation properties of the fermions.

The first two generations are chosen to have the same 
transformation properties as in the standard model,
\begin{eqnarray}
&&Q_L = (3,2,1)(1/3),\;\;\;\;U_R = (3,1,1)(4/3),\;\;\;\;D_R = (3,1,1)(-2/3),
\nonumber\\
&&L_L = (1,2,1)(-1),\;\;\;\;E_R = (1,1,1)(-2).
\end{eqnarray}

The numbers in the first parenthesis are the $SU(3)$, $SU(2)_L$
and $SU(2)_R$ group representations respectively, and the number in the second 
parenthesis is the $U(1)_{B-L}$ charge.

The third generation is chosen to transform differently,
\begin{eqnarray}
&&Q_L(3) = (3,2,1)(1/3),\;\;\;\;Q_R(3) = (3,1,2)(1/3),\nonumber\\
&&L_L(3) = (1,2,1)(-1),\;\;\;\;L_R = (1,1,2)(-1).
\end{eqnarray}

The above assignments are unusual compared with the conventional Left-Right
model, but they enhance the difference 
between the right handed couplings of the first two and the third generations.
This model is anomaly free.

The correct symmetry breaking and mass generation of particles 
can be induced by the vacuum expectation values of
three Higgs representations: 
$H_R = (1,1,2)(-1)$, which breaks the group down to $SU(3)\times SU(2)\times
U(1)$; and the two Higgs multiplets, 
$H_L = (1,2,1)(-1)$ and $\phi = (1,2,2)(0)$, which 
break the symmetry 
to $SU(3)\times U(1)_{em}$. For the purpose of symmetry breaking, only 
one of $H_L$ or $\phi$ is sufficient, but both are required 
to give masses to all fermions.
 
One may also introduce triplet Higgs multiplets, $\Delta_L = (1,3,1)(2)$
and $\Delta_R = (1,1,3)(2)$ to separate the $SU(2)_L$ and $SU(2)_R$
symmetry breaking scales and to give Majorana masses to the neutrinos. 
These triplets may be desirable for neutrino physics for example but they 
are not necessary for our present purposes.

The introduction of $\phi$ causes the standard model $W$ and $Z$ to mix 
with the new $W_R$ and $Z_R$ gauge bosons. Here $W_R$ is the $SU(2)_R$
charged gauge boson and $Z_R$ is a linear combination of
the neutral component of the $SU(2)_R$ gauge boson $W_{3R}$ and
the $U(1)_{B-L}$ gauge boson $B$ defined as

\begin{eqnarray}
Z_R = \cos\theta_R W_{3R} - \sin\theta_R B,
\end{eqnarray}
where $\tan \theta_R = g/g_R$.

In the bases $(W,\;W_R)$ and $(Z,\;Z_R)$ for the massive
gauge bosons, the mass matrices are given by,

\begin{eqnarray}
M^2_W = \left ( \begin{array}{ll}
m^2_{11W}&m^2_{12W}\\
m^2_{12W}&m^2_{22W}
\end{array}
\right ),\;\;\;\;
M^2_Z = \left ( \begin{array}{ll}
m^2_{11Z}&m^2_{12Z}\\
m^2_{12Z}&m^2_{22Z}
\end{array}
\right ).
\end{eqnarray}
with

\begin{eqnarray}
m^2_{11W} &=& 
{1\over 2}g^2_L (|v_L|^2 + 2 |v_{\Delta_L}|^2 + |v_1|^2 + |v_2|^2);
\nonumber\\
m^2_{22W} &=& {1\over 2} g_R^2
(|v_R|^2 + 2 |v_{\Delta_R}|^2 + |v_1|^2 + |v_2|^2);
\nonumber\\
m^2_{12W} &=& -g_Lg_R Re(v_1v_2^*),\nonumber\\
m^2_{11Z} &=& 
{1\over 2}{g^2_L\over \cos^2\theta_W} 
(|v_L|^2 + 4 |v_{\Delta_L}|^2 + |v_1|^2 + |v_2|^2);
\nonumber\\
m^2_{22Z} &=& {1\over 2} {g^2_R \over \cos^2\theta_R}
((|v_L|^2 + 4|v_{\Delta_L}|^2) \sin^4\theta_R
+ (|v_1|^2 + |v_2|^2) \cos^4\theta_R
\nonumber\\
&+& (|v_R|^2 + 4 |v_{\Delta_R}|^2));
\nonumber\\
m^2_{12Z} &=& {1\over 4}{g_Lg_R}{\sin\theta_R\over \cos\theta_W}
((  |v_L|^2 + 4|v_{\Delta_L}|^2 ) \tan\theta_R 
-(|v_1|^2+|v_2|^2)\cot\theta_R )),
\end{eqnarray}
where $v_i$ are the vevs of the Higgs representations $H_{L,R}$,
$\Delta_{L,R}$ and $\phi$. 

To compare the fermion-gauge-boson couplings that result in this 
model with those in the standard model, 
we find it convenient to introduce the following definitions for 
gauge mixing angles, 
\begin{eqnarray}
&&\tan\theta_W ={g_Y\over g_L},\;\;
g_Y = g \cos\theta_R = g_R \sin\theta_R,\;\;\tan\theta_L 
= {g\over g_L},\nonumber\\
&&\cos\theta_W = {\cos\theta_L\over \sqrt{1-\sin^2\theta_L \sin^2\theta_R}},
\;\;\sin\theta_W = {\sin\theta_L\cos\theta_R
\over \sqrt{1-\sin^2\theta_L \sin^2\theta_R}}.
\end{eqnarray}

After diagonalization of the mass-squared matrices, the lighter and heavier 
mass eigenstates $(Z^1,\;Z^2)$ and $(W^1,\;W^2)$
are given by
\begin{eqnarray}
\left ( \begin{array}{l}
W^1\\
W^2
\end{array}
\right ) 
= \left ( \begin{array}{ll}
\cos\xi_W&\sin\xi_W\\
-\sin\xi_W&\cos\xi_W
\end{array}
\right )
\left (\begin{array}{l}
W\\
W_R
\end{array}
\right ),\;\;\;\;
\left ( \begin{array}{l}
Z^1\\
Z^2
\end{array}
\right ) 
= \left ( \begin{array}{ll}
\cos\xi_Z&\sin\xi_Z\\
-\sin\xi_Z&\cos\xi_Z
\end{array}
\right )
\left (\begin{array}{l}
Z\\
Z_R
\end{array}
\right ),
\end{eqnarray}
where $\xi_{Z,W}$ are the mixing angles, 
\begin{eqnarray}
\tan(2\xi_{W,Z}) = {2 m^2_{12(W,Z)}\over m^2_{11(Z,W)} - m^2_{22(Z,W)}}.
\end{eqnarray}
In principle $\xi_Z$ and $\xi_W$ are related, and this can introduce 
severe constraints from processes such as $b \rightarrow s \gamma$. 
However, in general we find that the two can be quite different. 
For example, in the limit where $g << g_R$ we find
\begin{eqnarray}
\xi_W &\approx & {2 {\rm Re}(v_1 v_2^\star) \over v_R^2 + 2 v_{\Delta R}^2 +
v_1^2 + v_2^2}{\sin\theta_R \over \tan\theta_W} \nonumber \\
\xi_Z &\approx & {v_1^2+v_2^2 \over v_R^2 + 4 v_{\Delta R}^2+v_1^2+v_2^2}
\cos^3\theta_R{\sin\theta_R \over \sin\theta_W}.
\end{eqnarray}
This limit is of interest because it is the one required by the 
$A^b_{FB}$ data as we will see in the next section. 

These results show that it is possible to have the mixing in the 
neutral sector be larger than the mixing in the charged sector by 
taking $v_1$ much larger (or smaller) than $v_2$; or by giving them 
a large relative phase.

The vevs of $H_L$ and $\phi$ will generate all the masses and mixings for the 
quarks. They also provide masses and  mixings for leptons. Neutrinos
in this model can also receive Majorana masses from the vevs of $\Delta_L$
and $\Delta_R$. If $v_{\Delta_R}$ is much larger than the electroweak scale,
the right handed neutrino will be much heavier than the left-handed neutrinos.
However, there is also the possibility that the vev of $\Delta_R$ is of the
same order as the vev of $\Delta_L$ such that all neutrinos (the three 
left-handed ones and the right-handed one) are light. This possibility
may provide a solution to all the neutrino problems resulting from 
the atmospheric, solar and LSND data, should the LSND result be confirmed.

In this model there are new interactions between the massive 
gauge bosons and quarks. For the charged current interaction, there are
both left and right handed interactions. In the weak eigenstate basis, 
the charged gauge boson, $W$, couples to all generations, but 
the charged gauge boson, $W_R$, only couples to the third generation. 
There is a similar pattern for the neutral gauge interactions. This pattern 
gives rise to interactions between the fermions and the lightest 
physical gauge bosons that can be made to resemble the standard model 
couplings except for the right-handed couplings of the third generation; 
precisely the scenario suggested by the $A^b_{FB}$ data.
In the mass eigenstate basis the quark-gauge-boson interactions are 
given by,
\begin{eqnarray}
L_W&=& - {g_L\over \sqrt{2}} \bar U_L \gamma^\mu V_{KM} D_L
(\cos\xi_W W^{+1}_\mu - \sin\xi_W W^{+2}_\mu)\nonumber\\
&&-{g_R\over \sqrt{2}}
\bar u_{Ri} \gamma^\mu V^{u*}_{Rti}V^{d}_{Rbj} d_{Rj}
(\sin\xi_W W^{+1}_\mu + \cos\xi_W W^{+2}_{\mu}) ~+~{\rm h.~c.},
\end{eqnarray}
where $U = (u,\;\;c,\;\;t)$ and $D = (d,\;\;s,\;\;b)$. $V_{KM}$ is
the Kobayashi-Maskawa mixing matrix and $V^{u,d}_{Rij}$ are unitary matrices
which rotate the right handed quarks $u_{Ri}$ and $d_{Ri}$ 
from the weak eigenstate basis 
to the mass eigenstate basis. The repeated indices $i$ and $j$ are summed over
three generations. For the neutral sector the couplings are,
\begin{eqnarray}
L_Z &=& -{g_L\over 2 \cos\theta_W}
\bar q \gamma^\mu (g_V - g_A\gamma_5) q (\cos\xi_Z Z^1_\mu - \sin\xi_Z Z^2_\mu)
\nonumber\\
&+& {g_Y\over 2} \tan\theta_R  
({1\over 3} \bar q_L \gamma^\mu q_L+ {4\over 3} \bar u_{Ri} \gamma^\mu u_{Ri}
-{2\over 3} \bar d_{Ri}\gamma^\mu d_{Ri})
(\sin\xi_Z Z^1_\mu + \cos\xi_Z Z^2_\mu)\nonumber\\
&-& {g_Y\over 2} (\tan\theta_R + \cot\theta_R) (
\bar u_{Ri} \gamma^\mu V^{u*}_{Rti} V^{u}_{Rtj}u_{Rj} - 
\bar d_{Ri} \gamma^\mu V^{d*}_{Rbi} V^{d}_{Rbj} d_{Rj}) 
(\sin\xi_Z Z^1_\mu + \cos\xi_Z Z^2_\mu).
\label{neucoupl}
\end{eqnarray}
In this expression $q$ and $q_L$ are summed over  $u,d,c,s,t,b$ quarks, 
and repeated $i,j$ indices are summed over the three generations. 
The first line contains the standard model couplings to the $Z$ 
in the limit $\xi_Z =0$. The first two lines also contain couplings 
of the two $Z$ bosons to quarks that arise through mixing of the 
neutral gauge bosons. 

The most interesting terms occur in the third line and are 
potentially large if $\cot\theta_R$ is large. 
In the weak interaction basis they affect 
only the third generation whereas in the mass eigenstate basis 
(as written in Eq.~\ref{neucoupl}) they also give rise to flavor changing 
neutral currents. To satisfy the severe constraints that exist on 
flavor changing neutral currents we must have very small 
$V^d_{Rbd}$ and $V^d_{Rbs}$ matrix elements as we discuss in the next
section. 

It is clear that if $\xi_Z$ is not too small, through mixing, 
the $b$-quark coupling to the light Z boson can be very different from
that of the $d$ and $s$ quarks due to the last term in  Eq.~\ref{neucoupl}. 
If indeed 
the enhancement is achieved via a large value for $\cot\theta_R$, the 
couplings of the first two generations will remain close to their  
standard model values. This illustrates how this model would solve 
the $A^b_{FB}$ problem. To leading order in small $\xi_Z$ 
one finds,
\begin{eqnarray}
\delta g_{Lb} &\approx & -{1\over 6} \sin\theta_W \tan\theta_R \, \xi_Z
\nonumber \\
\delta g_{Rb} &\approx & {1\over 3}\sin\theta_W\tan\theta_R\, \xi_Z
- {1\over 2} \sin\theta_W (\tan\theta_R+\cot \theta_R) V^{d*}_{Rbb}V^d_{Rbb}
\, \xi_Z.
\label{maindev}
\end{eqnarray}
Similarly, one finds for the couplings of the top-quark to the $Z$ that, 
$\delta g_{Lt} = \delta g_{Lb}$, and
\begin{equation}
\delta g_{Rt} \approx  -{2\over 3}\sin\theta_W\tan\theta_R\, \xi_Z
+ {1\over 2} \sin\theta_W (\tan\theta_R+\cot \theta_R) V^{u*}_{Rtt}V^u_{Rtt}
\, \xi_Z.
\end{equation}

To explain the $A_{FB}^b$ anomaly the model must be able to generate 
the shifts of Eq.~\ref{fit}. The shift required for the left-handed 
coupling is small and at the level of radiative corrections. We 
have no way of fixing all the parameters of the model at one-loop so we 
concentrate on the larger shift required for the right-handed 
coupling. Keeping only the large $\cot\theta_R$ term Eq.~\ref{fit} implies that,
\begin{equation}
\xi_Z \cot\theta_R  V^{d*}_{Rbb}V^d_{Rbb} \sim 0.08.
\label{numanomaly}
\end{equation}
We now examine this result in view of the available phenomenological 
constraints. 

\section{Constraints}

As we have pointed out, the couplings of  Eq.~\ref{neucoupl} induce 
tree-level flavor changing neutral currents and there are severe 
phenomenological constraints on these. The potentially 
dangerous terms in Eq.~\ref{neucoupl} are of the form
\begin{equation}
{g_L \over 2}\tan\theta_W \cot\theta_R 
V^{q*}_{Rbi}V^q_{Rbj}   
\bar q_{Ri} \gamma^\mu q_{Rj} 
 Z^2_\mu.
\end{equation}
The easiest way to suppress this while keeping a large 
$Zbb$ right handed coupling is by choosing the $V^d_R$ matrix 
to be very close to the unit matrix. Usual constraints from $K-\bar K$, 
$D-\bar D$, and $B -\bar B$ mixing on four fermion operators such as 
those generated by a tree-level exchange of $Z_2$  imply that 
off-diagonal elements in $V^d_R$ and $V^u_R$ are of order $10^{-4}$ 
or less \cite{buch}. Since these matrices are arbitrary, 
choosing them to be approximately 
equal to the unit matrix does not constrain other sectors of our model.

In this model there are two mechanisms that generate a large 
contribution to the oblique parameter $T$ and this leads to 
constraints on the parameters that affect $A^b_{FB}$ in Eq.~\ref{numanomaly}.
First, there is a direct contribution to $T$ from $Z-Z_R$ 
mixing \cite{Holdom:1990xp} given by,
\begin{equation}
T = {1\over \alpha} \epsilon_1 = {1\over 
\alpha}\xi^2_Z \biggl({M^2_{Z_2} \over M^2_{Z_1}} -1
\biggr).
\label{tlimit}
\end{equation}
In addition there are large contributions from top (and bottom)-quark loops 
to the oblique corrections. Starting with the couplings in Eq.~\ref{maindev} 
these loop contributions can be obtained by extending the calculation 
of Ref.~\cite{Dawson:1995wa} to include the $Zbb$ couplings as well. 
Starting from the effective Lagrangian,
\begin{eqnarray}
{\cal L} &=& -{g \over \cos\theta_W} \sum_{q=t,b} 
\biggl((g_{Lq} + \delta g_{Lq}) 
\bar{q}_L\gamma_\mu q_L + (g_{Rq} + \delta g_{Rq}) 
\bar{q}_R\gamma_\mu q_R\biggr) Z^\mu \nonumber \\
&-& {g \over \sqrt{2}} \biggl[\biggl( (1+\delta\kappa_L)
\bar{t}_L \gamma_\mu b_L + \delta\kappa_R \bar{t}_R \gamma_\mu b_R\biggr)
W^{+\mu} +{\rm h.c.}\biggr],
\label{ancouplag}
\end{eqnarray}
one finds that the leading non-analytic contributions to the oblique 
parameters are \cite{Dawson:1995wa},
\begin{eqnarray}
S &=& {1\over 6\pi}\log\bigl({M^2_{Z_2}\over M_Z^2}\bigr)
\biggl(2(\delta g_{Rt} +\delta g_{Rb})- (\delta g_{Lt}+\delta g_{Lb})\biggr)
\nonumber \\
T &=& {3 \over 8 \pi \sin^2\theta_W}\biggl({M_t^2\over M_W^2}\biggr)
\log\bigl({M^2_{Z_2}\over M_Z^2}\bigr)
\biggl(2\delta\kappa_L+\delta g_{Rt} -\delta g_{Lt}\biggr)\nonumber \\
U &=& {1\over 2 \pi} \log\bigl({M^2_{Z_2}\over M_Z^2}\bigr)
\biggl(-4\delta\kappa_L+\delta g_{Lt}-\delta g_{Lb}\biggr).
\end{eqnarray}
The contributions to $U$ are seen to be small from Eq.~\ref{maindev}. 
There is a potentially large contribution to $S$ given by:
\begin{eqnarray}
S &=& {1\over 6\pi}\log\bigl({M^2_{Z_2}\over M_Z^2}\bigr)
\sin\theta_W \cot\theta_R \, \xi_Z V^{d*}_{Rbb}V^d_{Rbb} 
\biggl( {V^{u*}_{Rtt}V^u_{Rtt} \over V^{d*}_{Rbb}V^d_{Rbb}} -1\biggr)
\nonumber \\
&\sim & 0.007 
\biggl( {V^{u*}_{Rtt}V^u_{Rtt} \over V^{d*}_{Rbb}V^d_{Rbb}} -1\biggr). 
\label{sloop}
\end{eqnarray}
In the last line we have used Eq.~\ref{numanomaly} and taken 
$M_{Z_2} \sim 600$~GeV as a plausible upper bound (as we will see below).
From Ref.~\cite{Groom:in} we know that $S=-0.03 \pm 0.11 (-0.08)$, so 
new physics contributions to $S$ are constrained to be less than 0.22 
at the 2$\sigma$ level. We conclude that there are no significant constraints
on our model from $S$. 

\begin{figure}[!htb]
\begin{center}
\epsfxsize=16.5cm
\centerline{\epsffile{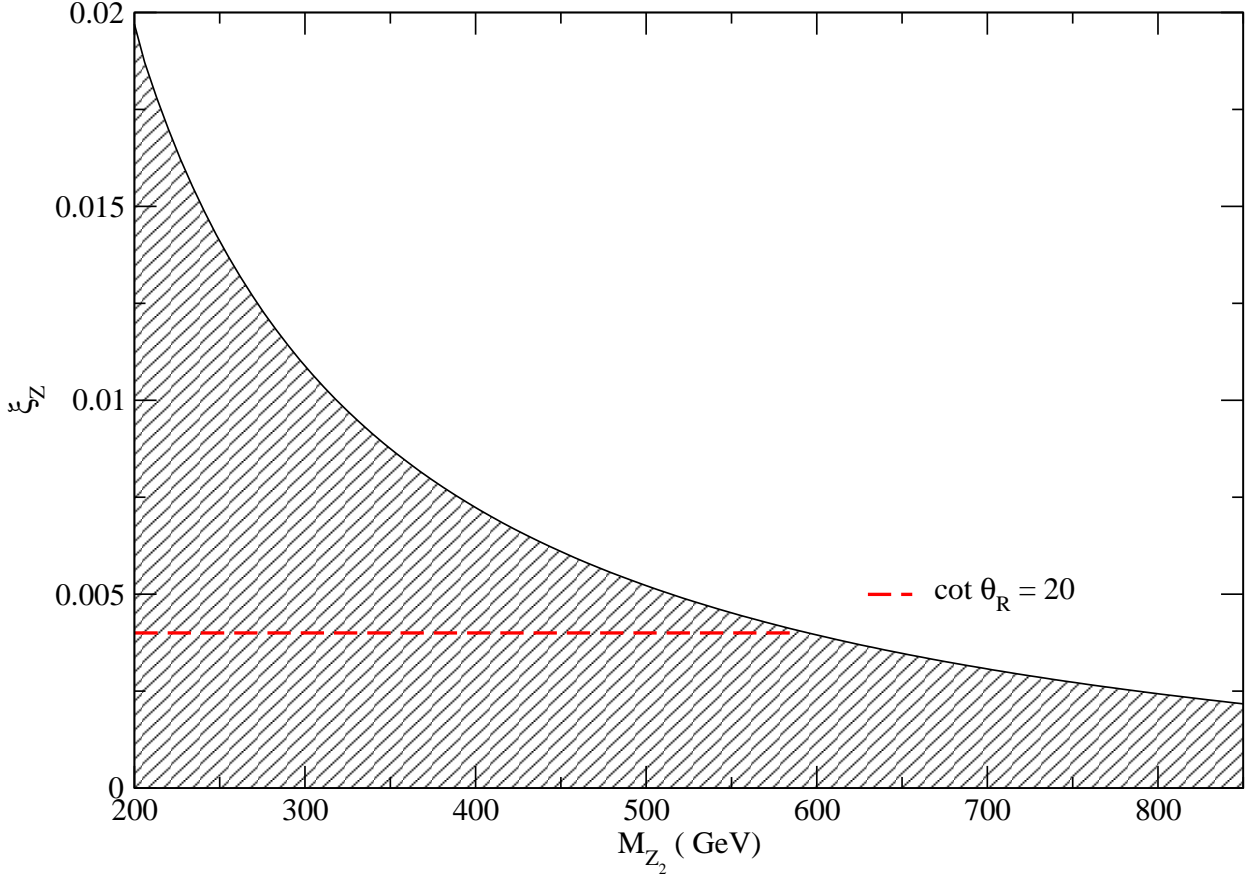}}
\end{center}
\caption{Allowed region (hatched) in $\xi_Z-M_{Z_2}$ from requiring $T<0.26$, 
a $2\sigma$ agreement with the global fit. Below the dashed line our 
model becomes non-perturbative as discussed in the text.}
\label{fig1}
\end{figure}

Returning to our discussion of $T$, we find a second large contribution,
\begin{equation}
T = {3 \over 16 \pi \sin\theta_W}\biggl({M_t^2\over M_W^2}\biggr)
\log\bigl({M_{Z_2}^2\over M_Z^2}\bigr)
\cot\theta_R \, \xi_Z V^{u*}_{Rtt}V^u_{Rtt}. 
\label{tloop}
\end{equation}
Combining Eqs.~\ref{tlimit} and ~\ref{tloop}, and using Eq.~\ref{numanomaly} 
restricts the allowed $\xi_Z - M_{Z_2}$ parameter space. In line with 
our discussion of flavor changing neutral currents we also require that 
$V^{u*}_{Rtt}V^u_{Rtt}/V^{d*}_{Rbb}V^d_{Rbb} \sim 1$.  The 
global fit from the WWW 2001 update to Ref.~\cite{Groom:in} is
\begin{eqnarray}
T = -0.02 \pm 0.13 (+0.09).
\end{eqnarray}
With the particle data book definition of $T$ this implies that new physics 
contributions to $T$ are at most $0.26$ at the 2$\sigma$ level, and we 
show the resulting constraints in Figure~\ref{fig1}. The hatched region in the 
figure indicates the parameter space allowed in our model. 

Additional contributions to the oblique parameters arise from the 
Higgs sector of our model. The model contains a SM-like Higgs boson 
from $H_L$ which contributes in a manner similar to that of the 
SM Higgs. The remainder of the scalar sector is largely unconstrained 
and we shall not discuss it further in this paper.

It is instructive to discuss existing constraints on LR models. 
An early comprehensive analysis of weak neutral current data 
\cite{Amaldi:1987fu} found that $|\xi_Z| \leq 0.05$ was typical for
left-right models. The equivalent constraint for our model would be 
weaker since our $Z_2$ couplings to the first two generations are 
much weaker than in the usual LR models. Nevertheless, $|\xi_Z| \leq 0.05$ 
is satisfied in the allowed region in Figure~\ref{fig1}. 
The best direct search limits for a $Z_2$ boson reported in 
Ref.~\cite{Cvetic:1995zs} come from CDF data \cite{Abe:1994ns,newexp} and 
for a LR $Z_2$ are of the order of $630$~GeV. However this limit 
assumes couplings of electroweak strength between the new $Z_2$ 
boson and the first two generations. In our model these couplings are 
at least ten times smaller (since they are proportional to $\tan\theta_R$ 
and we need $\cot\theta_R \geq 10$) effectively reducing the 
{\it lower bound} on the $Z_2$ mass to less than 100~GeV 
from experiments that only involve couplings to the first two generations. 
Ref.~\cite{Lynch:2000md} has studied the problem 
of placing bounds on the mass of a $Z_2$ that couples preferentially 
to the third generation. From searches for compositeness there are bounds 
on the scale of four fermion operators such as $\bar q_L \gamma_\mu q_L 
\bar e_L \gamma^\mu e_L$ on the order of $4$~TeV \cite{compo}. However, 
when these operators are induced by the exchange of the $Z_2$ in our model, 
the smallness of its couplings to the first two generations results 
in a very weak lower bound (well below 100~GeV) as well. 

It appears that there are no significant constraints on the mass of 
a $Z_2$ with couplings to the first two generations as weak as those in our 
model. For illustration we will simply assume a lower limit on 
the mass of our $W_2$ of order $100$~GeV because it has not been 
produced in $e^+e^-$ colliders, and assume a similar lower limit 
on the $Z_2$ mass to produce Figure~\ref{fig1}.

It is possible to place a theoretical constraint on the parameters 
of the model by requiring the coupling of the $Z_2$ to the third 
generation fermions to remain perturbative. Demanding that
\begin{equation}
\biggl({g_Y \cot\theta_R \over 2}\biggr)^2 \leq 4\pi
\end{equation}
results in $\cot\theta_R \leq 20$. Combined with Eq.~\ref{numanomaly},  
and assuming that $V^{d*}_{Rbb}V^d_{Rbb} \sim 1$, this results in 
a lower limit on $\xi_Z \geq 0.004$ shown as a dashed line in Figure~\ref{fig1}.
As seen in  Figure~\ref{fig1}, this also implies that $M_{Z_2}\leq 600$~GeV. 
The hatched region {\it below} the dashed line in  Figure~\ref{fig1} satisfies 
phenomenological constraints but implies a $Z_2$ coupling to the third 
generation which is non-perturbative.

We now turn our attention to the charged gauge boson sector. The 
early bounds on $W-W_R$ mixing from a comprehensive analysis of 
low energy data can be found in Ref.~\cite{Langacker:1989xa}. 
Depending on the model their typical result was,
\begin{equation}
|\xi_g| \leq 10^{-3} 
\end{equation}
where 
$\xi_g \equiv {g_R \over g_L} \, \xi_W = \tan\theta_W/\sin\theta_R\, \xi_W$. 

As with the neutral gauge boson sector, these constraints do not apply 
directly to our model. The best bound on $W_R$ couplings to 
third generation quarks comes from $b\rightarrow s \gamma$ as in the 
bound on the anomalous coupling $\delta \kappa_R$ from Eq.~\ref{ancouplag}
obtained in Ref.~\cite{fuji}. A more careful treatment of QCD corrections 
can be done along the lines of Ref.~\cite{Cho:zb}. The dominant 
contribution to $b\to s \gamma$ and the associated $b\to s g$ 
is from $W-W_R$ mixing, one has
\begin{eqnarray}
H_{\rm mixing} &=& 
-{4G_F\over \sqrt{2}} V_{tb}V_{ts}^* [c^{LR}_7 O_7 + c^{LR}_{8}O_8],
\nonumber\\ 
O_7 &=&{e\over 16\pi^2} m_b \bar s \sigma_{\mu\nu} R F^{\mu\nu}b,
\nonumber\\
O_8&=&{g_3\over 16\pi^2} m_b \bar s \sigma_{\mu\nu} R G^{\mu\nu} b,
\end{eqnarray}
where $c^{LR}_{7,8}$ are the Wilson Coefficients due to Left-Right mixing
evaluated at a scale of order $O(m_W)$. In our model they are given by
\begin{equation}
c^{LR}_7 = \xi_{eff}  {m_t\over m_b} \tilde F(x_t) \, ~,~~\, \,
c^{LR}_8 = \xi_{eff}  {m_t\over m_b} \tilde G(x_t)
\end{equation}
where,
\begin{eqnarray}
\xi_{eff} &=& {\tan\theta_W \over \sin\theta_R} \xi_W 
\biggl({V^u_{Rtt}V^{*d}_{Rbs} \over V^*_{ts}} + 
{V^{*u}_{Rtt}V^{d}_{Rbb} \over V_{tb}} \biggr),
\nonumber\\
\tilde F(x) &=& 
{-20+31 x -5x^2\over 12(x-1)^2} + {x(2-3x)\over 2(x-1)^3}\ln x,
\nonumber\\
\tilde G(x) &=& -{4+x+x^2\over 4 (x-1)^2} + {3x\over 2(x-1)^3} \ln x,
\end{eqnarray}
with $x_t = m^2_t/m_W^2$, and $V_{tb}$, $V_{ts}$ the usual CKM angles. 

Running down to the scale relevant to $B$ decays, we obtain the
dominant effective Wilson coefficient for $b\to s \gamma$, $c_{7eff}$,
\begin{eqnarray}
c_{7eff} = \eta^{16/23} c^{LR}_7 + {8\over 3} (\eta^{14/23} - \eta^{16/23})
c^{LR}_8.
\end{eqnarray}
Here $\eta = \alpha_s(m_{W})/\alpha_s(m_b)$.

Compared with the SM top quark contribution, there is an enhancement factor
$m_t/m_b$. Using the most recent experimental data for $b\to s\gamma$, 
$B(b\to s\gamma) = (3.21\pm 0.43\pm 0.27^{+0.18}_{-0.10})
\times 10^{-4}$\cite{cleo1} we find at the $2\sigma$ level that there are 
two allowed ranges for $\xi_{eff}$. They correspond to destructive and 
constructive interference with the standard model amplitude respectively 
and are,
\begin{eqnarray}
-0.032 < & \xi_{eff} & < -0.027 \nonumber \\
-0.0016 < & \xi_{eff} & < 0.0037
\label{ranxief}
\end{eqnarray}
In line with our discussion of flavor changing neutral currents we assume 
that the $V^{u,d}_R$ matrices are very close to the unit matrix. The largest 
contribution is then from 
\begin{equation}
\xi_{eff} \sim {\tan\theta_W \over \sin\theta_R}\, \xi_W
\end{equation}
With $\cot\theta_R \sim 10$ the two allowed ranges for $\xi_W$ are
\begin{eqnarray}
-0.006 < & \xi_W & < -0.005 \nonumber \\
-0.0003 < & \xi_W & < 0.0007.
\label{bsgrange}
\end{eqnarray}
For comparison, the bound on $\xi_Z$ from $T$ combined with the 
perturbative requirement resulted in $0.004< \xi_Z < 0.02$. 
In the first allowed range in Eq.~\ref{bsgrange} one finds 
$0.25 < |\xi_W/\xi_Z| < 1.5$ making both mixing 
angles of the same order. On the other hand, in the second allowed 
range in Eq.~\ref{bsgrange} $0 < |\xi_W/\xi_Z| < 0.18$, so $\xi_W$ 
is typically much smaller. An analysis of the quark mass matrices in our model 
reveals that it is possible to accommodate naturally a hierarchy 
$v_1/v_2 \sim m_t/m_b$.\footnote{Similar ratios 
of vevs arise naturally in $SO(10)$ Grand Unified Models for example 
\cite{kelley}.} This scenario would result in a $|\xi_W/\xi_Z|$ as 
low as $2 m_b/m_t \sim 0.046$. 

In conclusion 
we have shown that it is possible to accommodate the $A^b_{FB}$ result 
in a model with new right handed interactions for the third generation. 
The model predicts large deviations from the standard model in 
the right handed couplings of the top-quark and perhaps the tau-lepton. 
A striking feature of the 
model is the possible existence of a light ($M_{Z_2} < 600$~GeV) $Z^\prime$ 
boson on which there is no meaningful lower bound at present. The 
range allowed by $b\rightarrow s \gamma$ in Eq.~\ref{ranxief} also 
indicates that this model can give rise to CP asymmetries in B decays that 
deviate significantly from the standard model \cite{alaa}.

\noindent {\bf Acknowledgments}$\,$    
The work of X.G.H. 
was supported in part by National Science Council under grants NSC
89-2112-M-002-058, and in part by the Ministry of
Education Academic Excellence Project 89-N-FA01-1-4-3.
The work of G.V. was supported in part by DOE under 
contract number DE-FG02-01ER41155.

\end{document}